\documentclass[twocolumn,aps,10pt,prl,showpacs,showkeys,]{revtex4}
\usepackage[dvips]{graphicx}

\begin{document}
\title{The Jahn-Teller theorem for 3d ions and their compounds}
\author{R. J. Radwanski}
\affiliation{Center of Solid State Physics, S$^{nt}$Filip 5,
31-150 Krakow, Poland \\
Institute of Physics, Pedagogical University, 30-084 Krakow,
Poland}\homepage{http://www.css-physics.edu.pl}
\email{sfradwan@cyf-kr.edu.pl}
\author{Z. Ropka}
\affiliation{Center of Solid State Physics, S$^{nt}$Filip 5,
31-150 Krakow, Poland}

\begin{abstract}
In contrary to customarily consideration of the Jahn-Teller
theorem for 3d-ion compounds in the orbital space only we point
out that it has to be considered in the spin-orbital space. It is
despite the weakness of the intraatomic spin-orbit coupling. A
direct motivation for this paper is an erroneous claim of a
recent Phys. Rev. Lett. {\bf 96} (2006) 027201 paper that the
high-spin $t_{2g}^{4}e_{g}^{2}$ state of the 3$d^{6}$
configuration is not Jahn-Teller active.

\pacs{75.10.Dg : 71.70.-d} \keywords{Jahn-Teller theorem,
electronic structure, crystal field, spin-orbit coupling, 3d
magnetism}\vspace {-1.6 cm}

\end{abstract}
\maketitle \vspace {-1.6 cm} There is ongoing long-lasting
discussion in the magnetic community about a relationship between
the electronic structure, magnetism, the local 3$d$-ion
surrounding and a (off-cubic) lattice distortion in 3$d$ ionic
compounds like LaMnO$_3$, LaCoO$_3$ or YTiO$_{3}$. In this
discussion terms like the Jahn-Teller (J-T) effect and the J-T
distortions anavoidably appear which obviously are related to the
J-T theorem. This long-lasting theoretical discussion reveals
that the physical picture for 3$d$-ion compounds is far from being
established. On other side, experiments, e.g. X-ray-absorption
fine structure (XAFS) experiments provide more and more accurate
data on the local surrounding at the atomic scale which are
generally discussed in connection to the J-T effect.
\begin{figure}\vspace {+0.8 cm}
\begin{center}
\includegraphics[width = 8.7 cm]{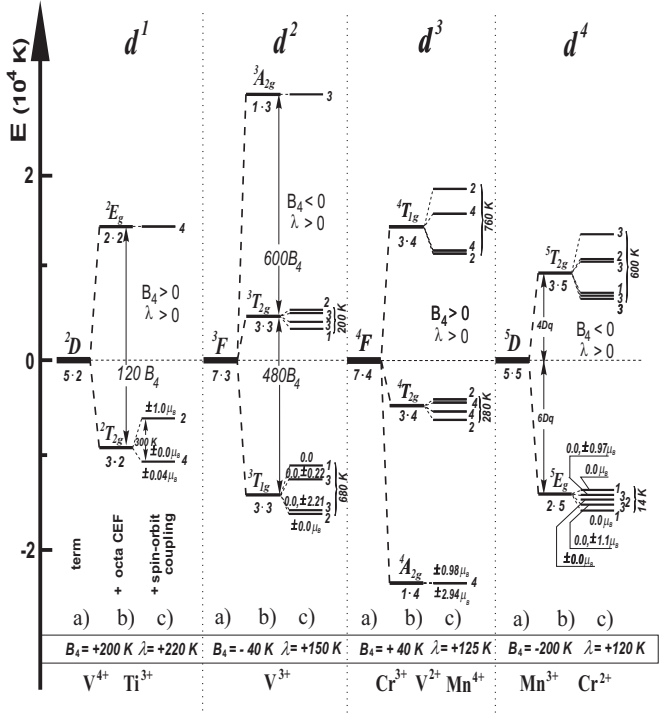}
\end{center} \vspace {-0.2 cm}
\caption{The calculated electronic structure of the 3$d^{n}$
configurations of the 3$d$ ions, 1$\leq n$ $\leq 4$, in the
octahedral crystal field (b) and in the presence of the spin-orbit
coupling (c). According to the Quantum Atomistic Solid-State
theory the atomic-like electronic structures, shown in (c), are
preserved also in a solid. (a) - shows the Hund's rule ground
term. Levels in (c) are labeled with degeneracies in the
spin-orbital space whereas in (b) the degeneracy is shown by the
orbital-spin degeneracy multiplication. The spin-orbit splittings
are drawn not in the energy scale which is relevant to CEF-only
levels shown in figures b. On the lowest localized level the
magnetic moment (in $\mu _{B}$~) is written. The shown states are
many electron states of the whole $d^n$ configuration. At zero
temperature only the lowest state is occupied. The higher states
become populated with the increasing temperature.}
\end{figure}
\begin{figure}[t!]\vspace {+0.8 cm}
\begin{center}
\includegraphics[width = 8.7 cm]{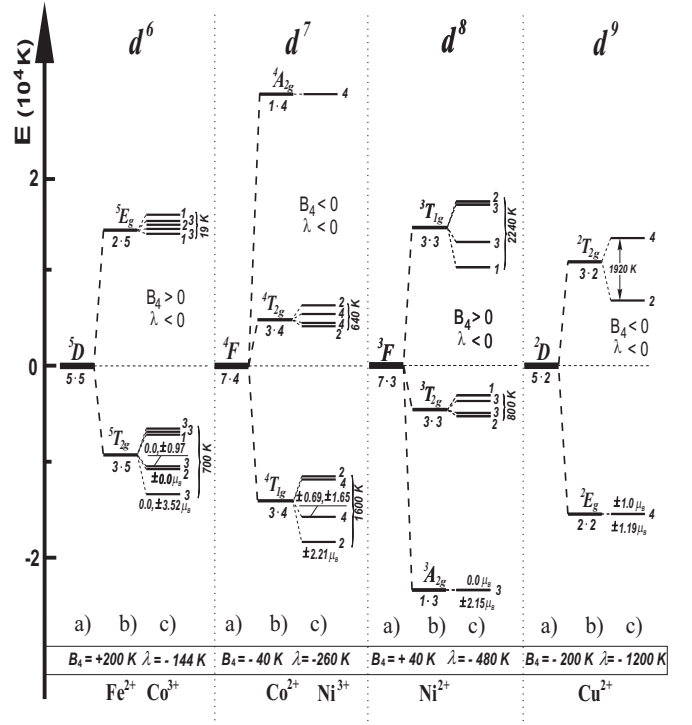}
\end{center} \vspace {-0.2 cm}
\caption{The calculated electronic structure of the 3d$^{n}$
configurations of the 3$d$ ions, 6$\leq n$ $\leq 9$, in the
octahedral crystal field in the presence of the spin-orbit
coupling. According to QUASST these atomic-like electronic
structures are preserved also in a solid.}
\end{figure}
The aim of this short Letter is to put an attention that the
physically adequate degeneracy space for the consideration of the
J-T theorem for the 3$d$ ions is the spin-orbital space. It is in
contrast to the current literature that considers the J-T
theorem, if applied to 3$d$ ions, in the orbital space only. A
direct motivation for this paper is an erroneous claim of a
recent Phys. Rev. Lett. {\bf 96} (2006) 027201 paper \cite {1}
that the high-spin (HS) state of the 3$d^{6}$ configuration is
not Jahn-Teller active. This erroneous conclusion was inferred by
consideration of the high-spin HS state as ascribed by
$t_{2g}^{4}e_{g}^{2}$ (S=2) configuration with two $e_{g}$
electrons. This situation was contrasting to the IS state, with
one $e_{g}$ electron following from a notation of the IS state as
$t_{2g}^{5}e_{g}^{1}$, as being J-T active, i.e. being a source
of lattice distortions.

We look on the J-T effect as related to a tendency for the
lowering energy of a system by the splitting and the lowering
energy of the predominantly lowest ionic state. This splitting is
an effect of lattice distortions or more generally of
lower-symmetry charge distribution.

We argue that the J-T theorem has to be considered in the
spin-orbital space. It is a direct consequence of the
intra-atomic spin-orbit coupling which, though weak in the 3$d$
ions, always exists. We are fully aware that we are not the first
considering the spin-orbital space and the spin-orbit coupling
but the appearance of this paper in Phys. Rev. Lett. proves the
need to remind this basic knowledge. Moreover, we would like to
put here attention to a very large degeneracy of 3$d$ states
being as is known from the atomic physics as ${10 \choose n} $. It
is 10, 45, 120 and 210 for n =1(9); 2(8), 3(7) and 4(6) d
electrons, respectively. It is in a sharp contrast to commonly
discussed 5 orbital states only. This large number of states in
the free ion is grouped in electronic terms. The ground state is
described by two Hund rules. The effect of the octahedral crystal
field has been studied by Tanabe and Sugano already 50 years ago
yielding well-known Tanabe-Sugano diagrams.

Figs 1c and 2c show the influence of the spin-orbit coupling on
the localized states of the strongly-correlated 3$d^n$ electron
system produced by crystal-field interactions of the octahedral
symmetry for the ground term described by two Hund's rules. It is
worth to note that in a solid a different ground term can be
realized. For instance, in LaCoO$_{3}$ the ground term is a
many-electron subterm $^{1}A_{1}$ originating from the $^{1}I$
term, which in the free Co$^{3+}$ ion lies 4.45 eV above the
ground term $^{5}D$ \cite {2}. The 13-fold degenerated $^{1}I$
term is split by the octahedral crystal field and the subterm
$^{1}A_{1}$ is strongly pushed down, by relatively strong crystal
field, due to its very large orbital quantum number $L$=6, as
occurs on the Tanabe-Sugano diagram for $Dq$/$B$=2.025. This
atomic-scale singlet $^{1}A_{1}$ subterm with $L$=6 is a low-spin
state $t_{2g}^{6}e_{g}^{0}$ ($S$=0) which can be written also as
$t_{2g}^{3}$$_{\uparrow}$$t_{2g}^{3}$$_{\downarrow}$.

Inspecting Figs 1c and 2c one can conclude that practically all
3$d$ ions are J-T ions, i.e. their electronic atomic-like
structure in the octahedral surrounding is very sensitive to the
detailed local symmetry. This prediction is in perfect agreement
with experimental observations - everywhere where exact
experimental studies are performed the octahedral surroundings is
found to be distorted. Even NiO is distorted despite the NaCl
structure and the orbital singlet ground state. An off-octahedral
distortion and magnetic interactions remove fully atomic-scale
degeneracy \cite {3}. The splitting of the 15-fold degenerated
$^{5}T_{2g}$ subterm by trigonal distortion in the presence of
the spin-orbit coupling has been presented in Ref. \cite{4} for
the Fe$^{2+}$ ion in FeBr$_{2}$ together with the formation of the
magnetic state below T$_{N}$ of 14 K.

The demanded spin-orbital space for the consideration of the J-T
theorem for 3$d$ ions is consistent with the standard treatment of
rare-earth-ion compounds.

In conclusion, we argue that the J-T theorem, if applied to
3$d$-ion compounds, has to be considered in the spin-orbital
space.

\end{document}